\documentclass[aps,prl,twocolumn]{revtex4}

\usepackage{subfigure}
\usepackage{graphicx}

\begin{document}

\vskip25pt

\title{Measurement of the Optical Absorption Spectra of Epitaxial Graphene from Terahertz to Visible}

\author{Jahan M. Dawlaty, Shriram Shivaraman, Jared Strait, Paul George, Mvs Chandrashekhar, Farhan Rana, and Michael G. Spencer}
\affiliation{School of Electrical and Computer Engineering, Cornell University, Ithaca, NY, 14853}
\author{Dmitry Veksler, Yunqing Chen}
\affiliation{Center for Terahertz Research, Rensselaer Polytechnic Institute, Troy, New York 12180}

\begin{abstract}
We present experimental results on the optical absorption spectra of epitaxial graphene from the visible to the terahertz (THz) frequency range. In the THz range, the absorption is dominated by intraband processes with a frequency dependence similar to the Drude model. In the near IR range, the absorption is due to interband processes and the measured optical conductivity is close to the theoretical value of $e^{2}/4\hbar$. We extract values for the carrier densities, the number of carbon atom layers, and the intraband scattering times from the measurements.
\end{abstract}

\maketitle

Graphene is a single atomic layer of carbon atoms forming a honeycomb crystal
lattice~\cite{dressel,nov0}. The unusual electronic and optical properties of
graphene have generated interest in both experimental and
theoretical arenas~\cite{nov0,nov1,NovoselovGeimFirst,zhang,heer}.
The high mobility of electrons in graphene has prompted a large
number of investigations into graphene based high speed electronic
devices, such as field-effect transistors and pn-junction diodes,
photonic devices, such as terahertz oscillators, and also low noise
electronic and optical sensors~\cite{NovoselovGeimFirst,lundstrom,marcus,gong,rana,
chemicalsensor}. For many of these applications, knowledge of the
optical properties of graphene is critical.

Graphene layers can be obtained via micromechanical
cleaving (exfoliation) of bulk graphite followed by careful
selection of monolayers by using optical, atomic force, or scanning
electron microscopes~\cite{nov1}. Although this technique results in
relatively high quality films, it might not be suitable for large
scale production. Recently, epitaxial growth of graphene by thermal
decomposition of SiC surface at high temperatures has been
demonstrated~\cite{heer,grapheneonSiC1}. This technique can provide
anywhere from a few monolayers of graphene to several ($>50$) layers
on the surface of a SiC wafer. Graphene layers grown by this
technique have demonstrated low temperature carrier mobilities
in the few tens of thousand cm$^{2}$/V-s range~\cite{heer}. In addition,
the electronic as well phononic properties of epitaxially grown graphene multilayers
have been found to be different from those of bulk graphite and
similar to those of a graphene monolayer~\cite{raman,stacking1,stacking2,stacking3}.
This observed difference in the properties of epitaxial graphene and bulk graphite
has been attributed to a different stacking scheme for carbon atom
layers in epitaxial graphene compared to that in bulk graphite in which the
layers are stacked according to the Bernal scheme~\cite{stacking1,stacking2,stacking3}.
Although the band energy dispersion in carbon atom multilayers can exhibit
massless Dirac Fermion like behavior at small energies for various
different stacking schemes and interlayer couplings, the
band energy dispersions at large energies can be significantly
different~\cite{stacking2,stacking3}. The exact structure of epitaxial graphene
and the nature of interlayer couplings remain active areas of investigation.
Measurement of the optical absorption spectra
over a wide frequency range can provide useful information about the structure
of epitaxial graphene.

In the visible to the mid-IR wavelength range ($\lambda<10$ $\mu$m), the optical absorption spectra of exfoliated graphene monolayers has been reported recently~\cite{stormer,geimopt,heinz}. In this paper, we report results from measurements of the optical absorption spectra of epitaxial graphene from the visible to the terahertz frequency range for the first time and compare the results with the theoretical predictions for graphene. In graphene, the valence and conduction bands resulting from the mixing of the $p_{z}$-orbitals are degenerate at the K (K') points of the Brillouin zone. Near these points, the tight-binding Hamiltonian in the nearest-neighbor approximation , using the basis consisting of orbitals centered on the $A$ and $B$ atoms, can be linearized and
written as~\cite{dressel},
\begin{equation}
H = \left[ \begin{array}{cc} \Delta & \hbar \, v \, (k_{x} + i\,k_{y}) \\
 \hbar \, v \, (k_{x} - i\,k_{y}) & - \Delta \end{array} \right] \label{eq0}
\end{equation}
where $v \approx 10^6$ m/s is the Fermi velocity. This Hamiltonian results in the energy dispersion relation for the conduction and
valence bands given by, $E_{\stackrel{C}{V}}(k) = \pm \sqrt{ \Delta^{2} +
(\hbar \, v \, k)^{2} }$. The bandgap is equal to $2\Delta$ and could acquire a non-zero value
as a result of any interaction that breaks the symmetry between the $A$
and $B$ atoms in the unit cell of graphene. Optical absorption in graphene
is described by the optical conductivity $\sigma(\omega)$. It can
be written as the sum of the interband conductivity $\sigma_{inter}(\omega)$ and
the intraband conductivity $\sigma_{intra}(\omega)$, both of which can be found using the Hamiltonian above and are given below~\cite{rana,sharapov}.
\begin{eqnarray}
\sigma_{inter}(\omega)& = & i \, \frac{e^2 \, \omega}{\pi}\int_{\Delta}^{\infty} d\epsilon \, \frac{\left(1+
\Delta^2/\epsilon^2 \right)}{(2\epsilon)^2-(\hbar \omega+i\Gamma)^2} \nonumber \\
& & \times \left[ f(\epsilon - E_{f}) - f(-\epsilon
- E_{f}) \right] \label{eq1}
\end{eqnarray}

\begin{eqnarray}
\sigma_{intra}(\omega) & = & i \, \frac{ e^2 / \pi \hbar^2}{ \omega +i/ \tau } \int_{\Delta}^{\infty}d\epsilon \,
\left(1+\Delta^2/\epsilon^2 \right) \nonumber \\
& & \times \left[ f(\epsilon-E_{f}) + f(\epsilon+E_{f}) \right] \label{eq2}
\end{eqnarray}
Here, $f(\epsilon - E_{f})$ is the Fermi distribution function with Fermi enery $E_{f}$, $\Gamma$ describes the
broadening of the interband transitions, and $\tau$ is the momentum relaxation time due to carrier intraband
scattering. The frequency dependencies of the real parts of $\sigma_{inter}(\omega)$ and $\sigma_{intra}(\omega)$
are depicted in Fig.1, assuming $\Delta=0$, $\Gamma=10$ meV and T=300K. Fig.1(a) shows the conductivities for $E_{f}=
-100$ meV and two different values of the scattering time $\tau$: 25 fs and 5 fs. Fig.1(b) shows the conductivities
for $\tau$ equal to 25 fs and two different values of the fermi energy $E_{f}$: 0 meV and -100 meV. At large
frequencies, the real part of $\sigma_{inter}(\omega)$ has a constant value equal to $e^{2}/4\hbar$. At small
frequencies, the real part of $\sigma_{inter}(\omega)$ approaches zero because interband optical transitions are
blocked due to the presence of electrons and holes near the band edges. The
plasmon dispersion and the free-carrier absorption in graphene are described by $\sigma_{intra}(\omega)$. Its frequency dependence is similar to that of a
Drude model, as it is evident from the pre-factor in Eq.(\ref{eq2}). The real part of $\sigma_{intra}(\omega)$
approaches zero for large frequencies at which carriers are unable to respond. At small frequencies, the real part
of $\sigma_{intra}(\omega)$ approaches the DC conductivity of graphene. Fig.1 (b) shows that the spectral shape of
$\sigma_{intra}(\omega)$ at small frequencies is strongly influenced by the intraband carrier scattering time
$\tau$. The total conductivity $\sigma(\omega)$ has a minimum in the frequency range where both
$\sigma_{intra}(\omega)$ and $\sigma_{inter}(\omega)$ are small.

\begin{figure}[tbp]
\centerline{
{\includegraphics[width=10.0cm]{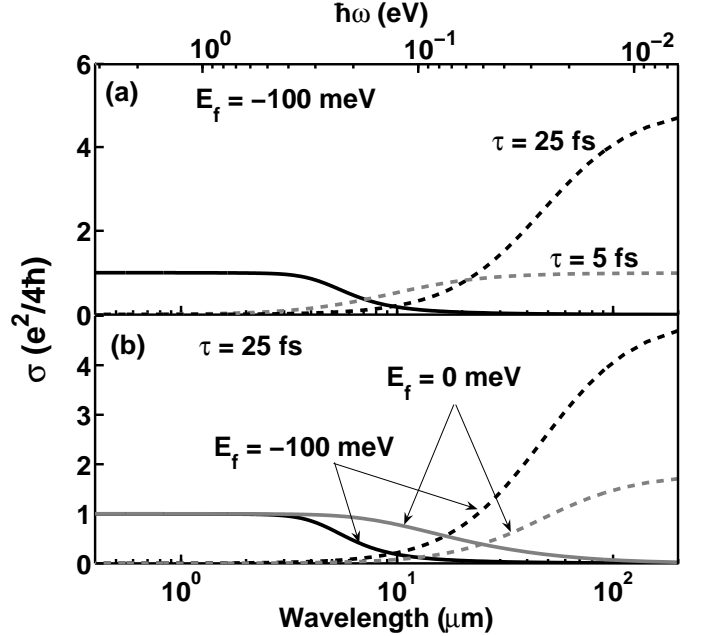}}} 
\caption{Real parts of the interband (solid) and intraband
(dashed) optical conductivities of graphene are plotted. (a) $E_{f}= -100$ meV and $\tau$ equals 25 fs and 5 fs. (b) $\tau = 25$ fs and $E_{f}$ equals 0 meV and -100 meV. Values of $\Gamma$ and $\Delta$ are assumed to be 10 meV and 0, respectively, and T=300K.} 
\label{fig1}
\end{figure}

The epitaxial graphene samples used in this work were grown on the carbon face of semi-insulating 6H-SiC wafers using the
techniques that have been reported in detail previously~\cite{grapheneonSiC1}. The samples were grown at
temperatures of $1400\,^{\circ}{\rm C}$ - $1600\,^{\circ}{\rm C}$ and pressures of $2-7\times10^{-6}$ torr. The number of carbon atom layers in each sample were estimated through X-ray photoelectron spectroscopy (XPS) using the Thickogram method~\cite{thicko}. Raman spectroscopy (using excitation wavelength of 488 nm) of the samples showed a single-resonant G peak close to 1580
cm${^{-1}}$, a double-resonant D' peak close to 2700 cm${^{-1}}$, and also a relatively low intensity
double-resonant D peak near 1350 cm$^{-1}$ \cite{raman,raman1}. The D peak is not allowed in perfect
graphene layers since it requires an elastic scattering process, which is made possible by disorder, to satisfy
momentum conservation~\cite{raman}. The presence of the D peak, therefore, indicates the presence of disorder in the
samples. The ratio of the intensities of the G and D peaks ($I_{G}/I_{D}$) has been shown to be
proportional to the crystal coherence length~\cite{ramandis}. The ratio $I_{G}/I_{D}$ for samples A, B and C were
$\sim$13, $\sim$17 and$\sim$2 respectively, indicating that sample C has a much larger level of disorder compared
to the other two samples. Ultrafast carrier dynamics in samples A and B have been studied in a previous work \cite{dawlaty}. Sample C reported in this paper is not the same as the one reported in our earlier study\cite{dawlaty}.

Three different instruments were used to measure the optical transmission through the graphene samples. In the
visible to the near-IR wavelength range (0.4-0.9 $\mu$m) a grating spectrometer was used. In the near-IR to the
mid-IR range (1.4-25 $\mu$m) a mid-IR Fourier Transform IR (FTIR) spectrometer was used. And in the mid-IR to the
far-IR (terahertz) wavelength range (15-200 $\mu$m) a far-IR FTIR spectrometer was used. The measured transmission
spectrum for each sample was normalized to the transmission spectrum of a reference SiC wafer. A small (few mm size)
aperture was used to ensure transmission through equal areas of the sample and the reference and also to avoid
corruption of data when changing instruments due to possible non-uniformities in the sample. The SiC substrate
transmits very little in the 6-14 $\mu$m wavelength range due to multi-phonon absorption~\cite{sicbands}. As a
result, the measured transmission spectra had poor signal-to-noise ratios in this wavelength range. The fringes
in the transmission spectra arising from multiple reflections within the SiC substrate were numerically filtered out
after normalizing with respect to the transmission spectrum of the reference SiC wafer.

\begin{figure}[tbp]
\centering
\subfigure{
\includegraphics[width=3.2in]{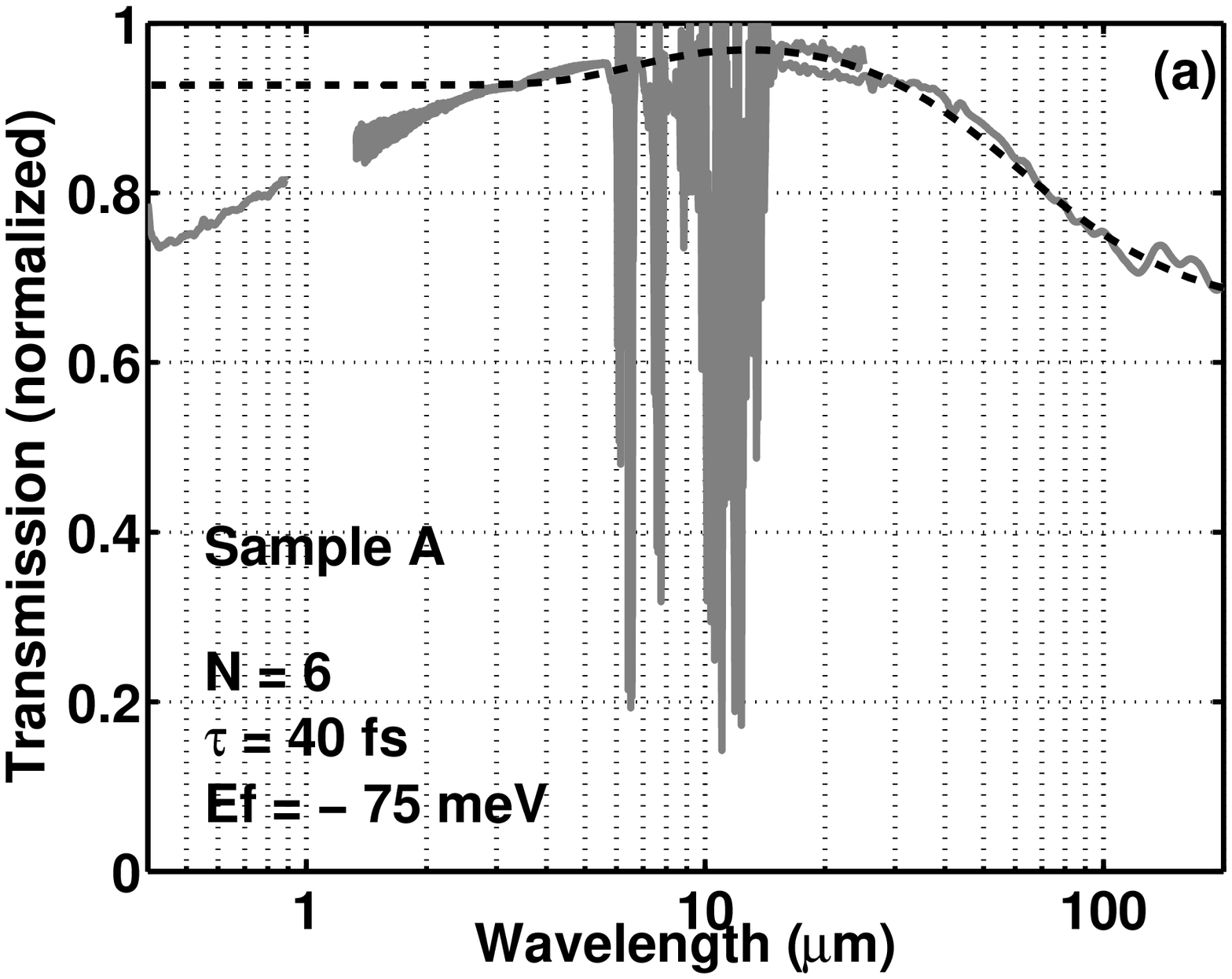}}
\vspace{0.0in}
\subfigure{
\includegraphics[width=3.2in]{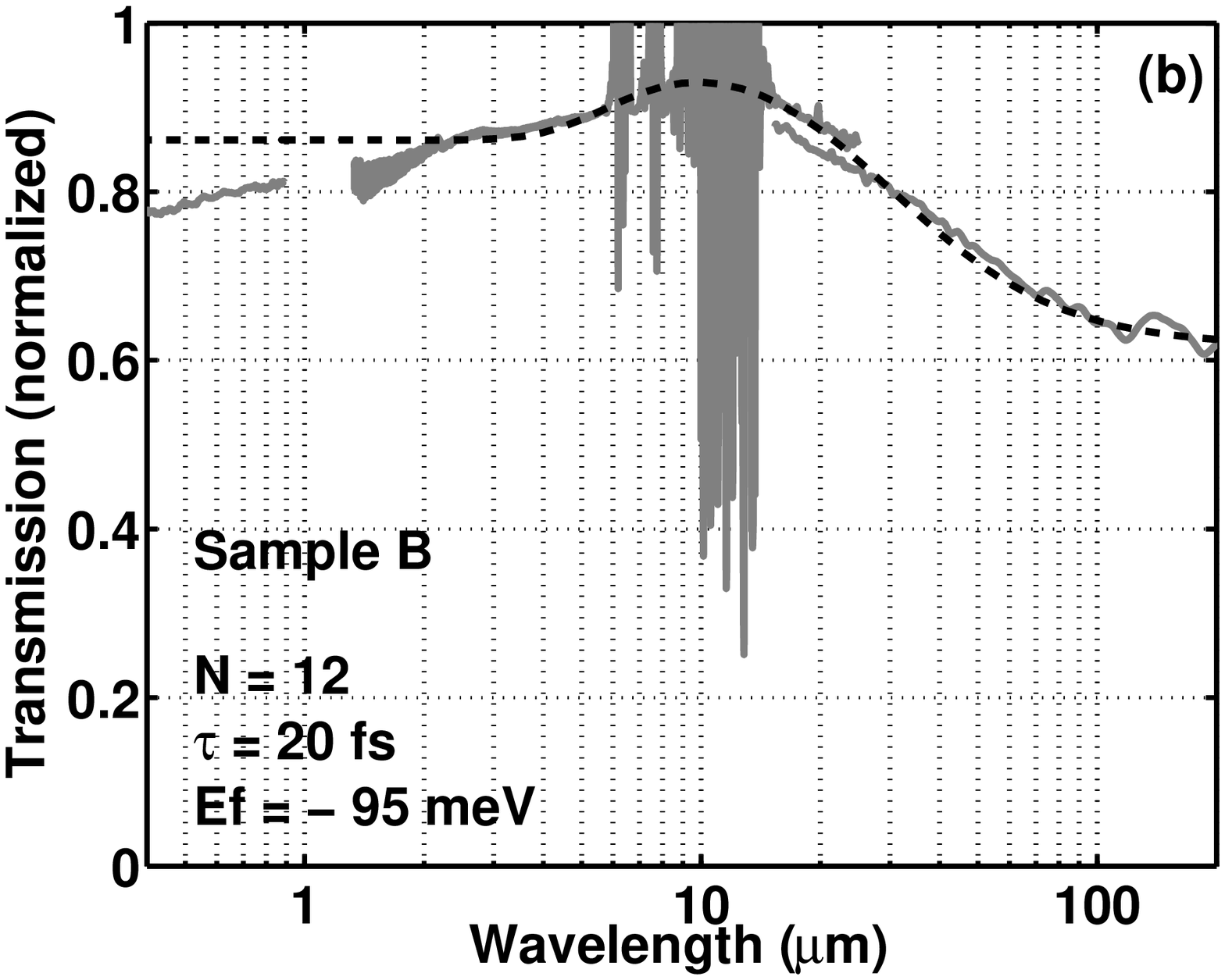}}
\vspace{0.0in}
\subfigure{
\includegraphics[width=3.2in]{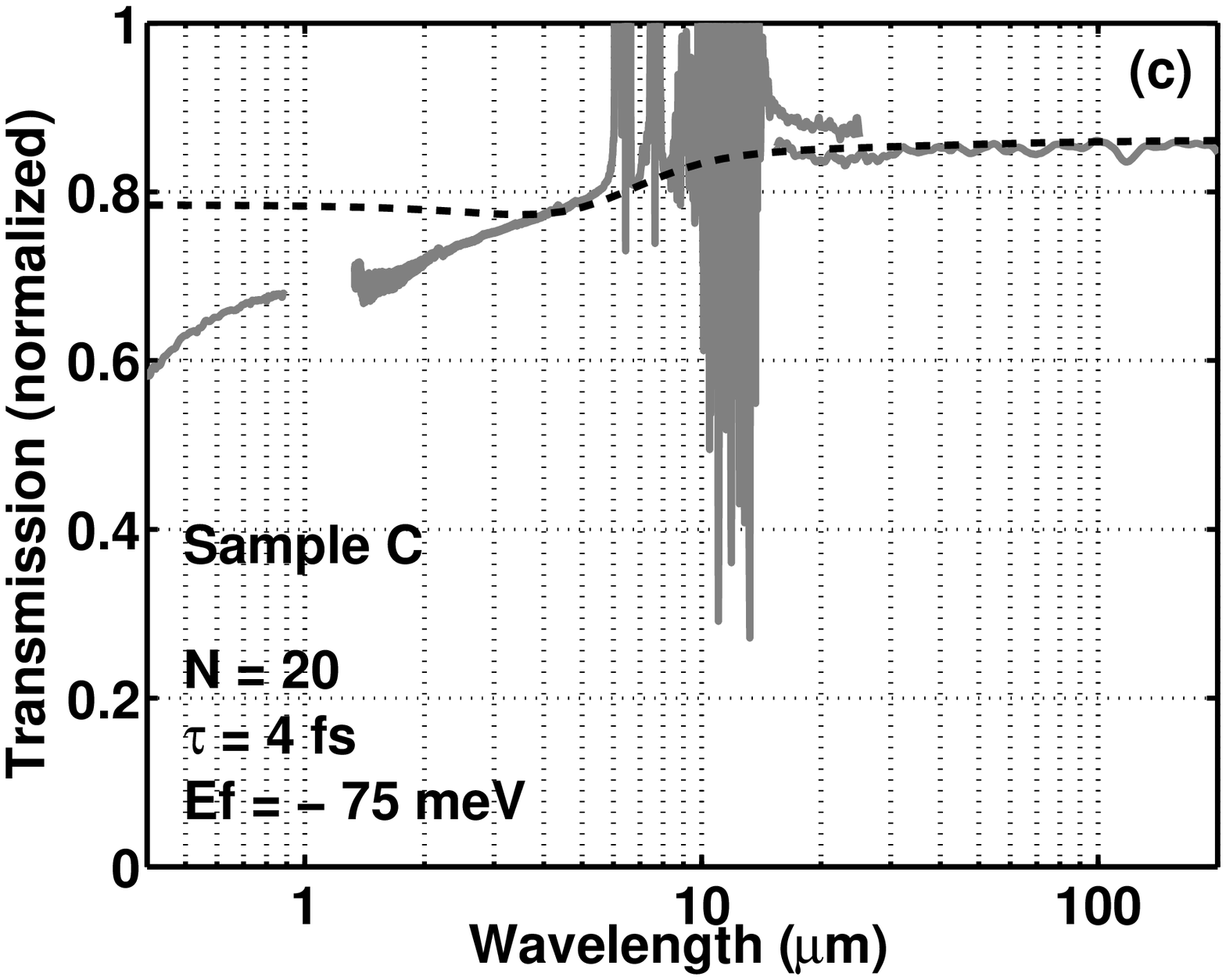}}
\caption{Measured transmission spectra (solid gray lines) of
samples A, B, and C from 0.4 $\mu$m to 200 $\mu$m along with the theoretical transmission spectra (dashed black line) using
Eqs.(\ref{eq1})-(\ref{eq3}). The values of the fitting parameters are included in the insets. The value of $\Delta$
is assumed to be zero.\label{fig2}}
\end{figure}

Fig.2 shows the normalized transmission spectra through the epitaxial graphene samples A, B and C. The optical
absorption in the graphene layers depends sensitively on the substrate index of refraction. Matching the optical
boundary conditions at the air/graphene/SiC interfaces, the optical transmission $T(\omega)$ through $N$ graphene
layers on a SiC wafer (normalized to the transmission through a plain SiC wafer) can be written in terms of the
complex optical conductivity $\sigma(\omega)$ as,
\begin{equation}
T(\omega) = | 1 +
N\sigma(\omega)\sqrt{\mu_{o}/\epsilon_{o}}/(1 +n_{SiC}) |^{-2}
\label{eq3}
\end{equation}
where $n_{SiC}\sim$~2.55 is the refractive index of SiC. For $N |\sigma(\omega)| \sqrt{\mu_{o}/\epsilon_{o}}/(1
+n_{SiC}) << 1$, $T(\omega)$ is related to only the real part of the optical conductivity and can therefore be used
to measure the absorption spectra of graphene. We have used Eqs.(\ref{eq1})-(\ref{eq3}) to model the measured
transmission data using $N$, $\tau$, $E_f$ as the fitting parameters. The parameters $\Gamma$ and $\Delta$ were assumed to be
$\sim$ 10 meV and $\sim$ 0 meV respectively. Changing the values of $\Gamma$ and $\Delta$ by small amounts (much less than $k_BT$) had little effect on the final results. The dashed black lines in Fig.2 are the theoretical fits to the experimental data (solid gray lines). The values of the fitting parameters are shown in the insets. A 5-10$\%$ variation in any one of the parameters, $N$, $\tau$,and $E_{f}$, produces a noticeable degradation in the quality of the theoretical fit to the data.

The number of graphene layers $N$ obtained this way agrees well with the value obtained through XPS. For example, the XPS method gave values of $N$ equal to 6 and 11 for samples A and B respectively. The extracted values of the Fermi energy correspond to average carrier densities of $\sim 5 \times 10^{11}$ cm$^{-2}$ and $\sim 8 \times 10^{11}$ cm$^{-2}$ per layer for samples A and B, respectively. Due to the electron-hole symmetry of the graphene bandstructure near the band edge, both negative and positive signs of the Fermi energy will fit the experimental data equally well. Also note that the experiments only give information on the total conductivity $N\sigma(\omega)$ of all the layers, and therefore the extracted value of the Fermi energy should be taken as an average value for all the layers. Recent work on epitaxial graphene suggests that a concentration of carriers larger than the intrinsic value is expected only in the first few carbon atom layers near the SiC interface. Assuming that only the first two layers have non-zero Fermi levels and the remaining layers are intrinsic, values of the Fermi level equal to -150 meV and -290 meV for the first two layers of samples A and B, respectively, provide good fits to the measured data. As mentioned earlier, Sample C is significantly more disordered than Samples A and B. The transmission spectra of Sample C shows a distinctly different shape in the THz region (Fig.2c), which can be fitted well with a very short carrier scattering time of $\sim$4 fs.

It has been recently pointed out that the first few graphene layers in epitaxially grown graphene could acquire a bandgap
as a result of interaction with the atoms in the SiC substrate~\cite{heer2}. Although a value of $\Delta$ equal to
zero was found to fit our measured data well, a value of $\Delta$ much smaller than $k_BT$ would be difficult to detect in our
measurements. A non-zero value of the bandgap $2\Delta$ in the mid-IR to far-IR range, where the intraband contribution to
the conductivity dominates, would have little effect on the transmission spectra. If the value of the bandgap
is in the near-IR to the mid-IR range, its effects at room temperature would be hard to distinguish from the reduction in
the interband conductivity at small frequencies due to band filling effects (see Fig.1(b)). Also, in multilayer
graphene structures the optical response is dominated by the large number of layers that are not close
to the substrate and do not have a bandgap.

The short wavelength end of the measured transmission spectra in Fig.2 deviates from the theoretical predictions for
wavelengths shorter than $\sim 2.5$ $\mu$m. The deviation is minimum for sample B and corresponds to $\sim$50$\%$ more
absorption at 0.4 $\mu$m compared to the theory. The reasons for this deviation are not clear. Two factors could be
responsible for this behavior: (i) the band energy dispersion and interband optical matrix elements at large
energies are different from those obtained from the Hamiltonian given in Eq.(\ref{eq1}), and (ii)
increased light scattering may be expected from the sample at shorter wavelengths as the wavelength approaches the
crystal coherence length ($\sim$50-100 nm).  Using a full-band tight-binding model for a graphene monolayer that
includes second and third neighbor interactions~\cite{tbaccurate}, and using the method for calculating the
interband optical matrix elements described by Johnson et.~al.~\cite{dressel2}, we obtain values of the optical
interband conductivity that deviate from $e^{2}/4\hbar$ only for wavelengths shorter than 1.5 $\mu$m and are only
$10\%-20\%$ larger than $e^{2}/4\hbar$ at wavelengths close to 0.4 $\mu$m. Therefore, effects related to trigonal warping and deviation of bands from linearity
cannot be completely responsible for the observed discrepancy. The band energy dispersion in epitaxial graphene at large energies could also be affected by the nature of the interlayer couplings. Note that the deviation
of the measured transmission spectra from the theory is not the same for the three samples indicating that disorder
might also have a role to play. More work is needed to investigate the nature of this discrepancy.

In conclusion, we have measured the optical absorption spectra of epitaxial graphene from the terahertz to the
visible frequencies. The experimental results have been shown to be in agreement with the theory except at short
wavelengths. Our results confirm the Drude-like frequency dependence of the intraband conductivity of graphene in the THz frequency range. The results presented here indicate that absorption spectroscopy can be used as a noninvasive technique to
characterize graphene films and find the values of parameters, such as the Fermi energy and the carrier density,
carrier intraband scattering time, and the number of graphene layers. The authors acknowledge support from the
National Science Foundation, the Air Force office of Scientific research contract No. FA9550-07-1-0332 (contract
monitor Dr. Donald Silversmith) and Cornell Material Science and Engineering Center (CCMR) program of the National
Science Foundation (cooperative agreement 0520404).

\end{document}